\documentclass[fleqn,multphys,vecphys]{svmult}

\usepackage{makeidx}   % allows index generation
\usepackage{graphicx}  % standard LaTeX graphics tool
\makeindex             % used for the subject index
\usepackage{hyperref}  % useful for online PDFs
\newcommand{\iDev}[1]{#1}
\newcommand{\iName}[1]{#1}

\newcommand{\iPostcode}[1]{#1}
\newcommand{\iCity}[1]{#1}
\newcommand{\iCountry}[1]{#1}
\newcommand{\email}[1]{\texttt{#1}}

\def\abstractname{Abstract.}%

\begin{document}

\title*{Raman imaging and electronic properties of graphene}          % title on the first page
\toctitle{Raman imaging and electronic properties of graphene}        % title as it will be printed in the contents
\titlerunning{Raman imaging and electronic properties of graphene}                                % title as it will be printed on the right-hand side of your paper

\author{F. Molitor, D. Graf, C. Stampfer, T. Ihn and K. Ensslin }                    % your name(s)
%\and Firstname2 Surname2 \inst{2}}
\authorrunning{F. Molitor, D. Graf, C. Stampfer, T. Ihn and K. Ensslin} % your names as it will be printed on the left-hand side of your paper's column line

\institute{\iDev{Laboratory for Solid State Physics},                  % Please use for dividing the parts of your address
\iName{ETH Zurich}, \newline
%\iStreet{Newtonstr. 15},
\iPostcode{8093}
\iCity{Zurich},
\iCountry{Switzerland}\newline
\email{ensslin@phys.ethz.ch}
%\and
%\iDev{Institut f\"ur Theoretische Physik II},
%\iName{Heinrich-Heine-Universit\"at D\"usseldorf},\newline
%\iStreet{Universit\"atsstra\ss e 1},
%\iPostcode{40225}
%\iCity{D\"usseldorf},
%\iCountry{Germany}\newline
%\email{author2@host.de}
}

\maketitle

\begin{abstract}
                Graphite is a well-studied material with known electronic and optical properties. Graphene, on  the other hand, which is just one layer of carbon atoms arranged in a hexagonal lattice, has been studied theoretically for quite some time but has only recently become accessible for experiments. Here we demonstrate how single- and multi-layer graphene can be unambiguously identified using Raman scattering. Furthermore, we use a scanning Raman set-up to image few-layer graphene flakes of various heights. In transport experiments we measure weak localization and conductance fluctuations in a graphene flake of about 7 monolayer thickness. We obtain a phase-coherence length of about 2 $\mu$m at a temperature of 2~K. Furthermore we investigate the conductivity through single-layer graphene flakes and the tuning of electron and hole densities via a back gate.
\end{abstract}

\section{Introduction}
The interest in graphite has been revived in the last two decades with the advent of fullerenes \cite{Kroto85} and carbon nanotubes \cite{Iijima91}. 
In a pioneering series of experiments it has become possible to transfer single- and few-layer graphene to a substrate \cite{Novoselov04}. Transport measurements revealed a highly-tunable two-dimensional electron/hole gas which mimics relativistic Dirac Fermions embedded in a solid-state environment \cite{Novoselov05b, Zhang05}. The term "relativistic" describes the linear energy-wave vector relation which gives graphene its exceptional electronic properties. Going to few-layer graphene, however, disturbs this unique system in such a way that the usual parabolic energy dispersion is recovered.
The large structural anisotropy makes few-layer graphene therefore a promising candidate to study the rich physics at the crossover from bulk to purely two-dimensional systems. Turning on the weak interlayer coupling while stacking a second layer onto a graphene sheet leads to a branching of the electronic bands and the phonon dispersion at the K point. Double-resonant Raman scattering \cite{Thomsen00} which depends on electronic and vibrational properties turns out to be an ingenious tool to probe the lifting of that specific degeneracy. Here we show scanning Raman images of graphene flakes and compare them with scanning force images. We evaluate the intensity, position and width of various Raman lines in order to quantify the numbers of monolayers in a given flake. Furthermore we determine the inelastic mean free path in a few-layer graphene wire and estimate the mobility in a single-layer graphene flake.

\section{Raman spectrum of graphene}
The energy gain and loss of scattered photons is related to the creation and annihilation of phonons at specific points in the phonon spectrum. For graphite the electron is excited from the valence band $\pi$ to the conduction band $\pi$* close to the K point in the electronic bandstructure. In graphene the unit cell is composed of two atoms leading to 6 phonon branches: Three acoustic branches starting at zero frequency and three optical branches at higher energies. The phonon spectrum of graphite has been calculated (see, e.g., \cite{Wirtz04}).

\begin{figure}
\includegraphics[width=10cm]{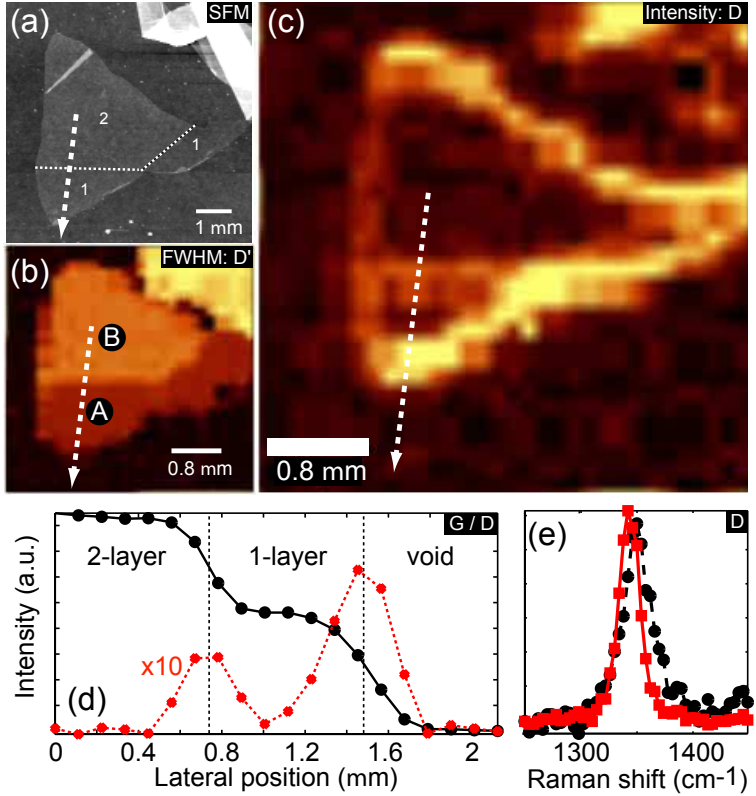}
    \caption{(a) SFM micrograph of a graphitic flake consisting of one double- and two single-layer sections (white dashed line along the boundaries), highlighted in the Raman map (b) showing the FWHM of the D' line.  (c) Raman mapping of the integrated intensity of the D line: A strong signal is detected along the edge of the flake and at the steps from double- to single-layer sections. (d) Raman cross section (white dashed arrow in (c)): Staircase behavior of the integrated intensity of the G peak (solid line) and pronounced peaks at the steps for the integrated intensity of the D line (dashed line). (e) Spatially averaged D peak for the crossover from double to single layer (disk, dashed line) and from single layer to the SiO$_2$ substrate (square, solid line). Taken from Ref. \cite{Graf07}.}
      \label{fig1}
\end{figure}

\begin{figure}
\includegraphics[width=8cm]{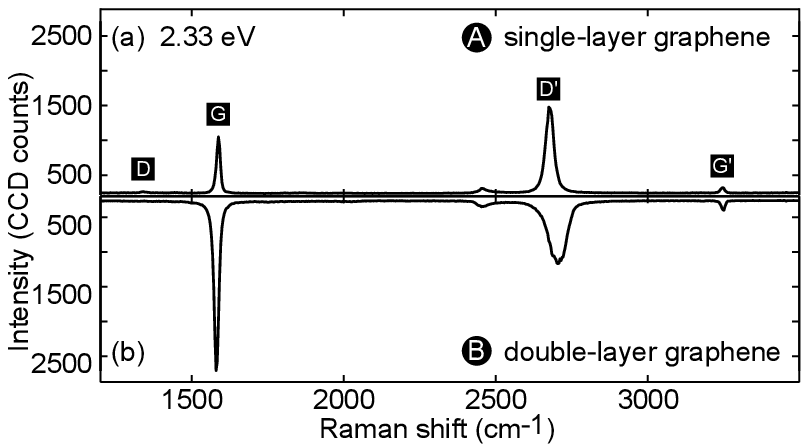}
    \caption{Raman spectra of (a) single- and (b) double-layer graphene (collected at spots A and B, see Fig. \ref{fig1}(b). Taken from Ref. \cite{Graf07}.}
      \label{fig2}
\end{figure}

In Fig.~\ref{fig1}(a) we start by presenting an image of a graphene flake taken with a scanning force microscope (SFM). The flakes were prepared following the method described in Ref. \cite{Novoselov04}. The number of monolayers is marked by a number in the respective flake area. Fig.~\ref{fig2} shows Raman spectra taken with a laser spot within one of the respective areas corresponding to one and two layers of graphene. The Raman spectrum of graphite has four prominent
peaks. For a recent review see Ref.~\cite{Reich04}. The G line is a standard Raman signal arising from the E$_{\mathrm{2g}}$ in-plane vibration of the atoms. In first approximation its intensity increases monotonously with the amount of material. The D line (D for defect) usually has a pronounced intensity if a material has defects \cite{Tuinstra70} or is strongly bend like in the case of carbon nanotubes. The absence of the D line in Fig.~\ref{fig2} is already a good indication for the structural quality of our graphene flakes. In the case of the D' line a second phonon excitation instead of an elastic backscattering is required \cite{Thomsen00}. The D' line is highly sensitive to the underlying energy dispersion. Consequently the D'-line is an ideal indicator to discriminate mono- from double layers in a given flake. The two monolayer areas show a single narrow peak, see Fig.~\ref{fig2}, while the double layer area shows a broadened peak with substructure. The width of the D' peak or - at high resolution - its splitting into different sub-peaks is  explained in the framework of the double-resonant Raman model \cite{Thomsen00}. A very similar analysis of the Raman lines of few-layer graphene was presented in Ref.~\cite{Ferrari06}.

\section{Raman imaging of graphene}
The most prominent difference in the spectra of single-layer, 
few-layer, and bulk graphite lies in the D' line:
the integrated intensity of the D' line stays almost constant, even though it narrows to a single peak at lower wave number at the crossover to a single layer (Fig.~\ref{fig2}). The width of the D' line is plotted in Fig.~\ref{fig1}(b) for the same area of the SFM scan in Fig.~\ref{fig1}(a). The outline of the flake and in particular the difference between the single and double layer areas can clearly be observed. We conclude that the width of the D' line is an excellent indicator to identify single-layer graphene. 

From cross-correlating the SFM micrograph in Fig.~\ref{fig1}(a) with the Raman map of the integrated D line (1300-1383 cm$^{-1}$) intensity in Fig. \ref{fig1}(c) we infer directly that the edges of the flake and also the borderline between sections of different height contribute to the D band signal whereas the inner parts of the flakes do not. This is somewhat surprising since for thinner flakes the influence of a nearby substrate on the structural quality should be increasingly important.
In the cross-section Fig.~\ref{fig1}(d) we see clearly that the D line intensity is maximal at the section boundaries, which can be assigned to translational symmetry breaking or to defects. However, we want to emphasize that the D line is still one order of magnitude smaller than the G line.
In Fig. \ref{fig1}(e) spatially averaged D mode spectra from the two steps shown in Fig.~\ref{fig1}(d) are presented. The frequency fits well into the linear dispersion relation of peak shift and laser excitation energy found in earlier experiments \cite{Vidano81}. In addition, we find that the peak is narrower and down-shifted at the edge of the single layer while it is somewhat broader and displays a shoulder at the crossover from the double to the single layer.

\section{Transport through few-layer graphene}
\begin{figure}
\includegraphics[width=10cm]{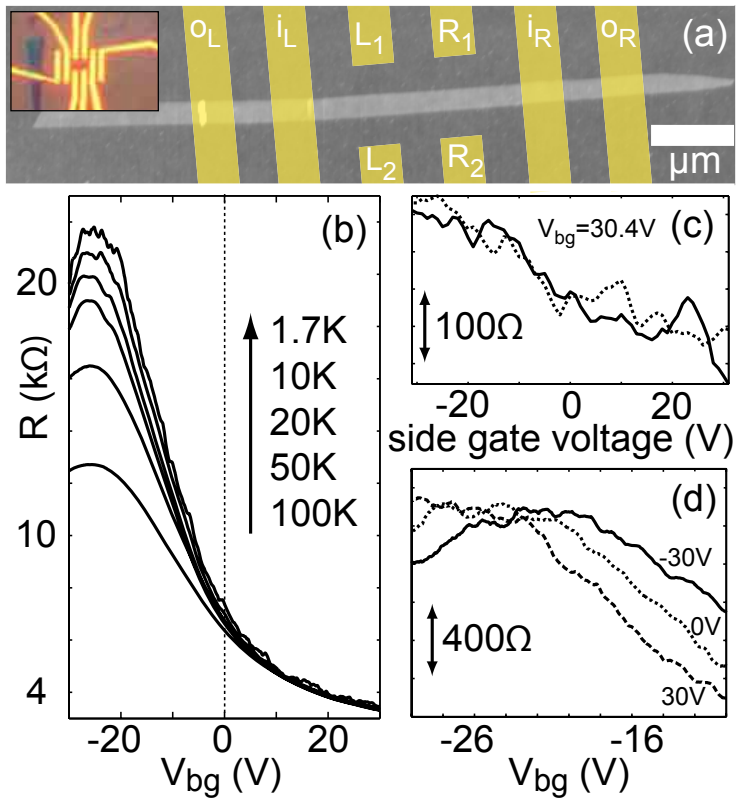}
    \caption{(a) SFM micrograph of a graphite wire resting ona silicon oxide surface with a schematic of the four contacts(iL, iR, oL, oR) and four side gates (L1, L2, R1, R2). Inset: Optical microscope image of the structure. (b) Four-terminal resistance as a function of back gate for different temperatures. (c) Resistance change as a function of the side gates L1+L2 (solid line) and R1+R2 (dotted line) at 1.7 K. (d) Resistance change as a function of back gate for different side gate voltages (L1+L2+R1+R2) at 1.7 K. Taken from Ref. \cite{Graf07b}.}
      \label{fig3}
\end{figure}

\begin{figure}
\includegraphics[width=8cm]{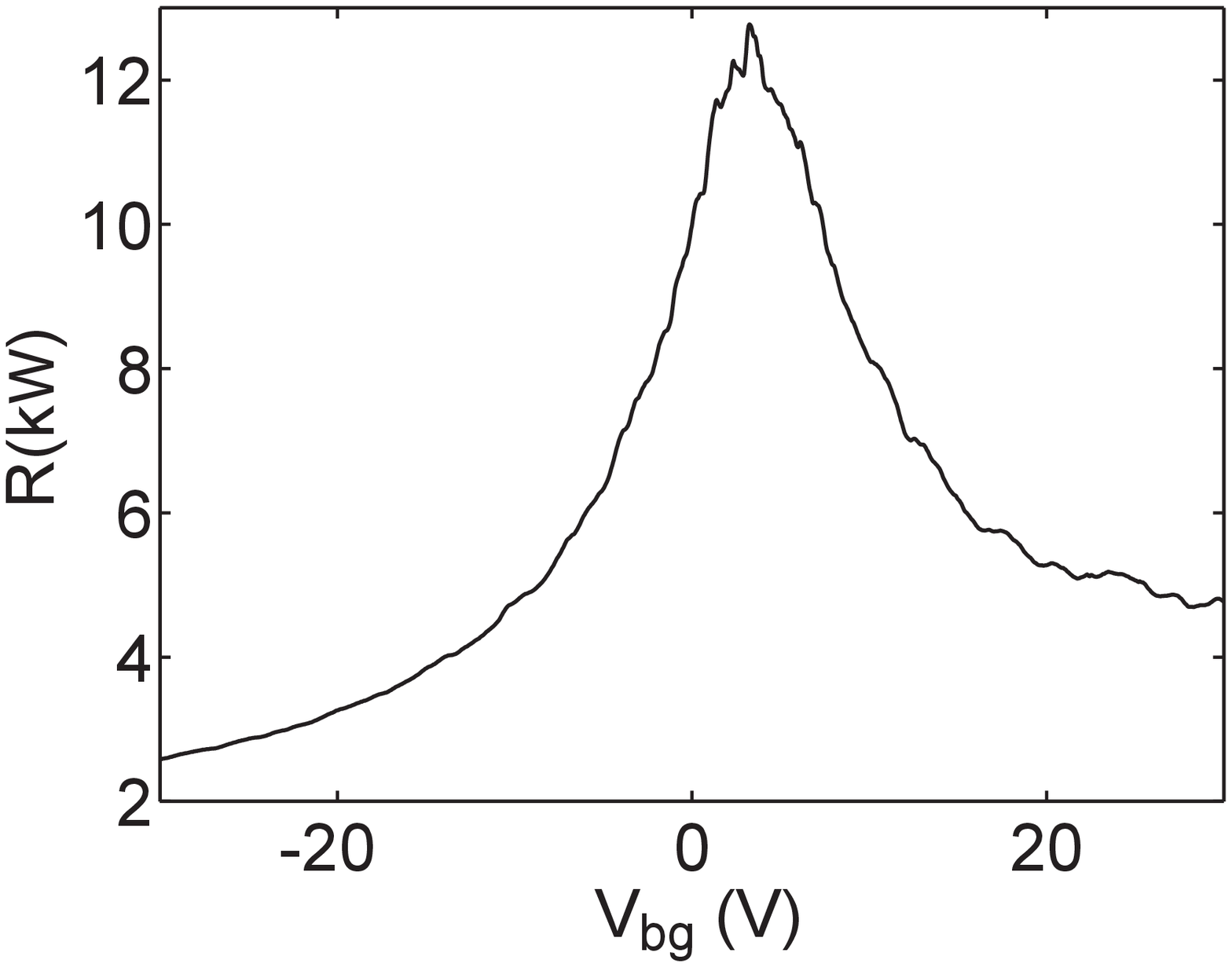}
    \caption{Two-terminal resistance of a single-layer graphene flake as a function of back gate voltage.}
      \label{fig4}
\end{figure}

Fig. \ref{fig3} shows transport measurements through a graphitic flake of several $\mu$m length, 320 nm in width and about 7 monolayers in height. Ohmic contacts and four in-plane gates as indicated by the gold colored areas were fabricated by electron beam lithography (SFM scan, Fig. \ref{fig3}(a)). We measured the resistance down to temperatures of 1.7 K as a function of back gate voltage (Fig. \ref{fig3}(b)) and side gate voltage (Fig. \ref{fig3}(c)) \cite{Graf07b}. All features in the resistance traces are reproducible. The metallic in-plane gates basically change the Fermi energy in a similar way as the homogeneous back gate, however, with a significantly reduced lever arm (Fig. \ref{fig3}(d)). We envision that in-plane gates could therefore serve well for the electrostatic tuning of nanostructures fabricated on graphene.

We also measured the resistance as a function of magnetic field and found well developed features corresponding to weak localization and conductance fluctuations \cite{Graf07b}. The data could be quantitatively analyzed in the framework of diffusive one-dimensional metals. We obtained a phase-coherence length of about 2 $\mu$m at a temperature of about 2 K. The temperature dependence is consistent with carrier-carrier scattering being the dominant dephasing mechanism.

Fig. \ref{fig4} shows the resistance of a single-layer graphene flake. The maximum resistance, i.e. the charge neutrality point, is close to zero back gate voltage indicating that doping is relatively weak. Using a plate-capacitor model we can infer the electron and hole densities for positive and negative gate voltages. We find a maximum mobility of about 7000 cm$^2$/Vs.
% If ever possible use BibTeX and the provided BibTeX-style for managing your references.
% Jabref a graphical  frontend is available on http://jabref.sourceforge.net/
% A short introduction is included in this package (ASSP_BibTeX_kurz_doku.pdf) (gernam only)
% If you use BibTeX please use the following two lines instedt of the thebibliography enviroment
%
% !!! Please include always your BibTeX database  !!!
%
%\bibliographystyle{assp}       % The BibTeX Style
%\bibliography{Chapter-1}      % Thea name of your BibTeX database

\section{Acknowlegdements}
We thank A. Jungen and C. Hierold for a fruitful collaboration during the measurement of the Raman spectra and L. Wirtz for theoretical support. Financial support from the Swiss National Science Foundation and NCCR Nanoscience is gratefully acknowledged.

\end{document}